\documentclass[12pt]{article}

 \newcommand{\be}{\begin{equation}}
 \newcommand{\ee}{\end{equation}}
 \newcommand{\ba}{\begin{eqnarray}}
 \newcommand{\ea}{\end{eqnarray}}
 \newcommand{\bl}{\begin{equation}\begin{array}{ll}}
 \newcommand{\el}{\end{array}\end{equation}}
 \newcommand{\bll}{\begin{equation}\begin{array}{lll}}
 \newcommand{\bdm}{\begin{displaymath}}
 \newcommand{\edm}{\end{displaymath}}
 \def\p{\partial}
 \def\f{\varphi}
 \def\ve{\varepsilon}
 
\def\lim{\rightarrow}
\def\half{\frac{1}{2}}

\def\dphi{\dot{\phi}}
\def\dvf{\dot{\varphi}}

\def\dF{\dot{F}}

\def\deta{\dot{\eta}}
\def\dxi{\dot{\xi}}

\def\be{\begin{equation}}
\def\ee{\end{equation}}
\def\bea{\begin{eqnarray}}
\def\eea{\end{eqnarray}}

\begin{document}

\title{Some Unusual Dimensional Reductions of Gravity: Geometric Potentials,
Separation of Variables, \\
and Static - Cosmological Duality}
\raggedbottom

\author{A.~T. FILIPPOV}

\maketitle

\begin{abstract}

We discuss some problems related to dimensional reductions of gravity
theories to two-dimensional and one-dimensional dilaton gravity models. We
first consider the most general cylindrical reductions of the
four-dimensional gravity and derive the corresponding (1+1)-dimensional
dilaton gravity, paying a special attention to the possibility of
producing nontrivial cosmological potentials from pure geometric variables
(so to speak, from `nothing'). Then we discuss further reductions of
two-dimensional theories to the dimension one by a general procedure of
separating the space and time variables. We illustrate this by the example
of the spherically reduced gravity coupled to scalar matter. This
procedure is more general than the usual `naive' reduction and apparently
more general than the reductions using group theoretical methods. We also
explain in more detail the earlier proposed `static-cosmological' duality
(SC-duality) and discuss some unusual cosmologies and static states which
can be obtained by using the method of separating the space and time
variables. This paper is a significantly extended and corrected sum of the
recent reports \cite{W1} and \cite{W2}.

\end{abstract}

\section{Introduction}
The procedure of dimensional reduction in classical physics is a well known
matter but, when one is working with gravity, some subtle points appear,
because geometric characteristics of the space-time become dynamical
variables. This is most obvious in the Kaluza - Mandel - Klein - Fock
reduction (KMKF reduction that usually but unjustly is called KK reduction),
in which the metric coefficients become physical fields.
This may look less clear in further reductions using
cylindrical or spherical symmetries. Then the effective space-time
becomes (1+1)-dimensional and some higher-dimensional metric coefficients
become dynamical fields, which mix with the original matter fields
produced by reductions from higher dimensions.

Low-dimensional models can be obtained by different chains of dimensional
reductions from higher-dimensional supergravity or gravity theories (see,
e.g., \cite{Julia} - \cite{VA}). For instance, we may consider toroidal
compactifications and KMKF reductions from eleven-dimensional theory to a
four-dimensional gravity coupled to Abelian gauge fields and scalar
fields. In case of spherical or cylindrical symmetry, we can further
reduce it to a one-dimensional dilaton gravity coupled to scalar matter
fields produced by the reductions. Roughly speaking, such a chain looks
like
\bdm
(1+10) \rightarrow (1+D) \rightarrow (1+3) \rightarrow (1+1) \, ({\rm  spherical
\,\,\, or \,\,\, cylindrical)} .
\edm
The two-dimensional theories describe inhomogeneous cosmologies, evolution of black holes,
and various types of waves (spherical, cylindrical, and plane waves).
Their further reductions give both the standard (or generalized) cosmological models and static
states (in particular, static black holes):
\bdm
(1+1) \rightarrow (1+0) \, ({\rm cosmological}) \,\,\,\, {\rm or} \,\,\,
(1+1) \rightarrow (0+1) \, ({\rm static}) .
\edm
It is also useful to keep in mind static chains:
\bdm
(1+3) \rightarrow (0+3) \, ({\rm general \,\, static}) \, \rightarrow (0+2) \,
({\rm axial})  \rightarrow (0+1) .
\edm
We do not consider here some other reductions, like the general
axial reduction
\bdm
(1+3) \rightarrow (1+2) \, ({\rm axial})  \, \rightarrow (1+1) \, ({\rm  spherical
\,\,\, or \,\,\, cylindrical)} \,\,\,  {\rm or \,\, (0+2) \,\, (axial)}  .
\edm
Note also that it is not necessary to use
step-by-step reductions. For instance, the (1+0)-dimensional homogeneous isotropic
cosmologies and (0+1)-dimensional static black holes
are usually derived by direct symmetry reductions from higher dimensions.

This is quite legitimate, if you are not interested in relations between
these reductions and are not trying to immerse them in a more general
formulation allowing for their dynamical treatment. In addition, when you
have many matter fields, considering first the (1+1)-dimensional dilaton
gravity allows us to obtain other interesting solutions, through which the
static states, cosmologies, and waves may be inter-related (about
relations between various types of solutions see \cite{VA}-\cite{VA1}).
Not less important is the fact that lower-dimensional dilaton gravity
theories may be regarded as Lagrangian or Hamiltonian systems that are
often integrable (in some sense) and thus we may hope to study them in
detail and even quantize them in spite of the fact that the general
quantum solutions of the higher-dimensional theories can not be
constructed. If we make the reductions with due care, we may possibly find
important information about solutions of higher-dimensional theories. To
succeed in this, one should follow a few important rules which must be
used in the process of dimensional reductions.

First, one should not make `excessive' gauge fixings before writing all
the equations of motion. For example, the number of independent fields in
the reduced theory must be not less than the number of the independent
Einstein equations for the Ricci tensor plus the number of the equations
for the matter fields. Otherwise, some solutions of the reduced theory
will  not satisfy (and often do not satisfy) the higher-dimensional
equations of motion. Second, by analogy with the usual (`naive')
reduction, it may be tempting to make all the fields to depend only on one
variable (space or time, if we consider reducing (1+1)-dimensional dilaton
gravity). By doing so one can loose some solutions that can be restored
with the aid of more general dimensional reductions (e.g., by separating
variables). We would like to also emphasize that the concept of
dimensional reductions should be understood in a broader sense. An example
of a more general dimensional reduction is given in \cite{A3}, \cite{VA1}:
the solutions of a (1+1)-dimensional integrable model depend on arbitrary
moduli functions of one variable; if these functions reduce to constants,
we obtain essentially one-dimensional solutions. This reduction may be
called a `dynamical dimensional reduction' or a `moduli space reduction'.
In this example, a class of reduced solutions of the two-dimensional
theory consists of those that essentially depend on two space-time
variables (say, $t$ and $r$), which nevertheless should be regarded as
`one-dimensional solutions' in a well defined but somewhat unusual sense.
Unfortunately, at the moment we can introduce this new dimensional
reduction only for explicitly integrable dilaton gravity theories.

To avoid misunderstanding, let us formulate the practical `philosophy'
behind our approach to the dimensional reduction of gravity. As distinct
from the common tendency to concentrate on geometric and symmetry
properties, we follow the Arnowitt - Deser - Misner approach to the
treatment of gravity by using Lagrangian and Hamiltonian dynamics with
constraints. Thus, the `geometric' variables are treated on the same
footing with other dynamical variables and the aim is not only to derive
the metric and other geometric properties of the space-time but to
construct Lagrangians and Hamiltonians and then to solve the dynamical
equations which, eventually, should be quantized. To quantize such a
complex nonlinear theory as gravity one should first find some simple
explicitly integrable approximation, like the oscillator approximation in
the standard QFT. Natural candidates for such `gravitational oscillators'
may be static states (e.g., black holes), cosmological models and some
simple gravitational waves. One may argue that all these objects are
somehow related to the Liouville equation rather than to the oscillator
equation \cite{A1}-\cite{VA1}.

Although one should not expect that such simplified models can
give completely realistic description of gravity, cosmology, or gravitational waves,
they may serve as a tool for developing a new intuition, which is so needed for
understanding new data on the structure of our Universe. They can also give reasonable
first approximations for constructing more realistic solutions as well as
some hints of how our main gravitational objects
are related physically (at the moment we find only mathematical relations).
Using explicitly integrable
models one can clearly see a duality between black holes and cosmologies as well as
observe that they both are limiting cases of certain gravitational waves. The duality
can also be seen in nonintegrable models (e.g., when we use a separation of variables),
while the `triality' including some gravitational
waves was up to now observed only in integrable theories \cite{A3} - \cite{VA1}.

The content of this paper is the following. In Section~2 we summarize the
main properties of rather general (1+1)-dimensional dilaton gravity models
describing dimensionally reduced (super)gravity theories. Section~3 deals
with the dilaton gravity theory obtained by the most general cylindrical
reduction of the four-dimensional gravity coupled to scalar fields. This
dilaton gravity is more general than usually considered and it was first
introduced in \cite{W2}. We show that the most general cylindrical
reduction gives an additional (`geometric') potential, which drastically
changes the properties of the popular integrable $SL_2/SO_2$~
$\sigma$-model dilaton gravity. In this paper we also discuss in more
detail further dimensional reductions of this generalized `cylindrical'
dilaton gravity. In Section~4 we apply the method of `dividing and
separating' introduced in \cite{W1} to the general `spherical' dilaton
gravity and reproduce the concept of duality between the static (in
particular, black holes) and cosmological solutions\footnote{ A relation
between the black holes and some cosmologies is known for long time, but
it was not clearly formulated and seriously investigated. Similar
relations between static states in gravity with matter were demonstrated
by some exact solutions of the spherical gravity coupled to matter
\cite{A1}. The necessity of generalizing the standard naive reductions was
explicitly demonstrated in \cite{VA}. }. We derive the general constraints
that must be satisfied to make the separation possible and briefly outline
the construction of several static and cosmological solutions (leaving the
complete classification to a future publication). In Conclusion we compare
the dimensional reductions considered here to those of papers \cite{VA} -
\cite{VA1}, in which we observed not only the static-cosmological duality
of the exact analytical solutions but also a relation of the static and
cosmological states to waves. We also outline the possibility of applying
the method of separating to cylindrical and axial static models.

\section{(1+1)-Dimensional Dilaton Gravity}
It is well known that there exist (1+1)-dimensional dilaton gravity theories
coupled to scalar matter fields, which are reliable
models for some aspects of high-dimensional black holes,
cosmological models, and branes.
The connection between high and low dimensions has
been demonstrated in different contexts of gravity and string
theory and in some cases allowed one to find general solution
or some special classes of solutions in high-dimensional
theories. In this paper, we only discuss reductions of the four-dimensional
gravity theory coupled to scalar fields. In fact, after reducing
to the dimension (1+1) all the matter fields are essentially equivalent
to the scalar ones.

For example, spherically symmetric gravity coupled to Abelian gauge
fields and massless scalar matter fields exactly reduces to a (1+1)-dimensional
dilaton gravity coupled to scalar fields
and can be explicitly solved if the
scalar fields are constants independent of coordinates.
Such solutions may describe interesting physical objects --
spherical static black holes, simplest cosmologies, etc.
However, when the scalar matter fields, which presumably play a
significant cosmological role, are not constant, few exact analytical
solutions of high-dimensional theories are known.
Correspondingly, the generic two-dimensional models of
dilaton gravity nontrivially coupled to scalar matter are usually not
integrable.

Some other important four-dimensional space-times, having
symmetries defined by two commuting Killing vectors, may also be described
by two-dimensional dilaton gravity. For example, the simplest Einstein - Rosen
cylindrical gravitational waves \cite{Einstein}
are described by a (1+1)-dimensional dilaton gravity coupled
to one scalar field. The simplest stationary axially symmetric
pure gravity \cite{Ernst} may be described
by a (0+2)-dimensional dilaton gravity coupled to one scalar field (this may be
related to the previous cylindrical case by the analytic continuation of one space
variable to imaginary values). Cylindrical waves attracted attention of many researchers
for many years (see, e.g., \cite{Kuhar}, \cite{Chandra1}). More recently,
similar but more general dilaton gravity models
were also obtained in string theory. Some of them may be solved by using
modern mathematical methods developed in the soliton theory (see, e.g., \cite{BZ} -
\cite{Alekseev}).

Let us briefly remind a fairly general formulation of the (1+1)-dimensional
dilaton gravity.
The effective Lagrangian of the (1+1)-dimensional dilaton gravity
coupled to scalar fields $\psi_n$, which can be
obtained by-dimensional reductions of a higher-dimensional spherically
symmetric (super)gravity, may usually be (locally) transformed to
the following form:
\bea
L = \sqrt{-g} \, [ U(\f) R(g) + V(\f,\psi) +
W(\f)(\nabla \f)^2    +
 \sum_n Z_{nm} \nabla \psi_n \nabla \psi_m ]. \,
\label{eq:1}
\eea
Here $g_{ij}(x^0,x^1)$ is a generic (1+1)-dimensional metric with signature (-1,1),
$g \equiv {\rm det}|g_{ij}|$ and $R \equiv R(g)$ is the Ricci curvature
of the two-dimensional space-time,
\be
ds^2=g_{ij}\, dx^i \, dx^j \, , \,\,\,\,\,\, (i,j = 0,1).
\label{eq:2}
\ee
The effective potentials $V$ and $Z_{nm}$
depend on the dilaton $\f (x^0,x^1)$ and on $N$ scalar fields
$\psi_n(x^0,x^1)$ \footnote{The potentials $Z_{nm}$ define a negative definite quadratic form.
}.
They may depend on other parameters characterizing the parent higher-dimensional
theory, e.g., on charges introduced by solving the equations
for the Abelian gauge fields, etc. There are two important simple cases:
1.~$Z_{nm}(\f, \psi) = \delta_{nm} Z_n(\f)$, and 2.~constant $Z_n$, independent of the fields.
The dilaton function $U(\f)$ is usually monotonic and one can put
(at least locally) $U(\f) = \f$ or $U(\f) = \exp ({-2\f})$, etc.
We also may use in Eq.~(\ref{eq:1}) a Weyl transformation
to exclude the gradient term for the dilaton, i.e. to make $W \equiv 0$.
Under the transformations to this frame (we may call it the Weyl frame)
the metric and the potential transform as
\be
g_{ij} \rightarrow {\tilde{g}}_{ij} \equiv w(\f) g_{ij} \, , \,\,\,\,
V \rightarrow \tilde{V} \equiv V/ w(\f) \, , \,\,\,\,
Z \rightarrow \tilde{Z} \equiv Z ,
\label{eq:2a}
\ee
where $w(\f)$ is defined by the equation
$w^{\prime}(\f) / w(\f) = W(\f) / U^{\prime}(\f)$.

As we mentioned above, in two-dimensional space-times all matter fields can
eventually be reduced to different scalar fields although, for keeping traces of
different symmetries,
it may be convenient to retain gauge fields, spinor fields, etc.
The Lagrangian (\ref{eq:1}) should be considered as an effective
Lagrangian. In general, it is equivalent to the original one on the
`mass shell' but the solutions of the original equations may be completely
recovered and used to construct the solutions of the higher-dimensional
`parent' theory. For a detailed motivation and specific examples see \cite{VA},
where references to other related papers can be found.

To simplify derivations we will use the equations of motion in the
light-cone metric, $ds^2 = -4f(u, v) \, du \, dv$ and
with $U(\f) \equiv \f$, $Z_{nm} = \delta_{nm} Z_n$,
$W \equiv 0$.
By first varying the Lagrangian in generic coordinates and then
going to the light-cone ones we obtain the equations of motion
 \be
 \p_u \p_v \f+f\, V(\f,\psi)=0, \label{eq:3}
 \ee
  \be
  f \p_i ({{\p_i \f} / f }) \, = \sum Z_n (\p_i \psi_n)^2\, , \,\,\,\,\,\,\,\,\,
 (i=u,v) \ .
 \label{eq:4}
 \ee
 \be
\p_v (Z_n \p_u \psi_n) +\p_u (Z_n \p_v \psi_n) + f V_{\psi_n}(\f,\psi)=
\sum_m Z_{m, \psi_n} \, \p_u \psi_m \, \p_v \psi_m \ ,
 \label{eq:5}
 \ee
 \be
 \p_u\p_v\ln |f| + f V_{\f}(\f,\psi) = \sum Z_{n, \f} \,\p_u \psi_n \, \p_v\psi_n \ ,
 \label{eq:6}
 \ee
 where $V_{\f}= \p_{\f} V$, $V_{\psi_n} = \p_{\psi_n} V$,
 $Z_{n, \f}= \p_{\f} Z_n$, and $Z_{m, \, \psi_n} =\p_{\psi_n} Z_m$.
These equations  are not independent. Actually,
(\ref{eq:6}) follows from (\ref{eq:3}) $-$  (\ref{eq:5}). Alternatively,
if  (\ref{eq:3}), (\ref{eq:4}), (\ref{eq:6}) are satisfied,
one of the equations (\ref{eq:5})
is also satisfied.

If the Lagrangian (\ref{eq:1}) was obtained by a consistent reduction of some
high-dimensional theory (i.e. not using gauge fixings, which reduce the number of
independent equations, and not applying non-invertible transformations to the coordinates
or unknown functions), the solutions of these equations can be reinterpreted as
special solutions of the parent higher-dimensional equations.

If the scalar fields are constant, $\psi = \psi_0$, these equations can be solved
with practically arbitrary potential $V$ that should satisfy only one
condition: $V_{\psi}(\varphi,\psi_0) = 0$, see Eq.(\ref{eq:5}).
 The constraints (\ref{eq:4}) then can be solved because their right-hand
sides are identically zero. It is a simple exercise to prove that there exist chiral
fields $a(u)$ and $b(v)$ such that $\varphi (u,v) \equiv \varphi (\tau)$
and $f(u,v) \equiv \varphi^{\prime} (\tau) \, a'(u) \, b'(v)$,
    where $\tau \equiv a(u) + b(v)$
(the primes denote derivatives with respect to the corresponding
argument). Using this result it is easy to prove that (\ref{eq:3})
has the integral $ \varphi^{\prime} + N(\varphi) = M $,
where $N(\varphi)$ is defined by the equation $N^{\prime}(\f) = V(\f , \psi_0)$
and $M$ is the constant (integral) of motion.
The horizon, defined as a zero
of the metric $h(\tau)\equiv M - N(\varphi)$, exists because the equation
$M = N(\varphi)$ has at least
one solution in some interval of values of $M$. These solutions are actually
one-dimensional (`automatically' dimensionally reduced) and can be interpreted
as black holes (Schwarzschild, Reissner Nordstr{\o}m, etc.)
or as cosmological models.

These facts are known for a long time and were derived
by many authors using different approaches. A similar solution was obtained in
the two-dimensional gravity with torsion \cite{MK1}. In the standard dilaton gravity,
first studied in detail in Ref.~\cite{Banks},
the local integral of motion $M$ was constructed in \cite{Geg} and, by a much
simpler derivation, in \cite{A1}. The equivalence of the two-dimensional gravity
with torsion to the standard dilaton gravity was shown in \cite{MK2}. The global solutions
of the pure dilaton gravity were constructed by many authors (see, e.g., \cite{Strobl}).
Systematic studies of matter coupled dilaton gravity models initiated by the CGHS
`string inspired' dilaton gravity model \cite{CGHS} resulted in finding more general
but simple enough integrable theories (see, e.g., \cite{A1}-\cite{A3}).
A review of different aspects of dilaton gravity and further references can also be found
in \cite{Strobl-r}, \cite{Kummer}.

With the pure dilaton gravity in mind, it looks, at first sight, natural to introduce
the following reduction of the two-dimensional dilaton gravity theories
to one-dimensional ones: let $\f$ and $\psi$ depend only on $\tau \equiv
a(u) + b(v)$, where $\tau$ may be interpreted either as the space or the time variable.
Then we obtain both the (0+1)-dimensional theory of static distributions of
the scalar matter (including black holes) and (1+0)-dimensional cosmological models.
However, analyzing their solutions (see simple examples in \cite{A1})
one can find that not all standard Friedmann cosmologies may be obtained in this way
\cite{VA}, \cite{A3}.
In view of the symmetry (`duality') between the (1+0) and (0+1)-dimensional
reductions one may conclude that not all static solutions are obtained by the naive reduction.
In other words, this simple (naive) procedure of dimensional reduction is not complete!
The same conclusion can be made if we use the space-time variables $(t,r)$.
Before discussing this phenomenon, we consider another simple source of a further
incompleteness in the standard processes of reductions.

\section{Generalized Cylindrical Reductions}
The last remark in the previous section signals that we should apply more care when
using dimensional reductions in gravity. To illustrate how more general reductions
may emerge we first discuss cylindrically symmetric reductions in the (1+3)-dimensional
pure gravity.
For acquiring a feeling of connections between the two-dimensional Lagrangian (\ref{eq:1})
and higher-dimensional theories let us consider
the four-dimensional cylindrically symmetric gravity coupled to one scalar
field:
\be
S_4 = \int d^4 x \, \sqrt{-g_4} \, [ R_4 + V_4 (\psi) + Z_4 (\psi) (\nabla \psi )^2] .
\label{eq:7}
\ee
Here the most general cylindrically symmetric metric should be used.
It can be derived by applying the general KMKF reduction. The corresponding metric
may be written as ($i,j = 0,1$; $m,n = 2,3$)
\be
ds_4^2 = (g_{ij} + h_{mn} A_i^m A_j^n) dx^i dx^j + 2 A_{im} dx^i dy^m +
h_{mn} dy^m dy^n \, ,
\label{eq:8a}
\ee
where all the metric coefficients depend only on the $x$-coordinates
($t,r$) while $y^m =(\phi, z)$ are some coordinates
on the two-dimensional cylinder (torus).

Usually, in the four-dimensional reduction the coordinate functions
$A_i^m$ are supposed to vanish \cite{Schmutzer},  but we will see in a moment
that this drastically changes the resulting two-dimensional dilaton gravity
theory. To see this, we also suppose that $\psi$ depends only on $x$ and
integrate out of Eq.(\ref{eq:8a})  the dependence on $y$. Extracting the dilaton
from the cylinder metric by writing
\be
h_{mn} \equiv \f \sigma_{mn} \, , \,\,\,\, \det(\sigma_{mn}) = 1 ,
\label{eq:9a}
\ee
and neglecting an inessential numeric factor, we find the two-dimensional
Lagrangian (in what follows we will omit the $V_4$ and $Z_4$ terms):
\be
L = \sqrt{-g} \, \{ \f [R(g) + V_4 + Z_4 (\nabla \psi )^2] + {1 \over 2\f} (\nabla \f)^2  -
{\f \over 4} {\rm tr} (\nabla \sigma \sigma^{-1}  \nabla \sigma \sigma^{-1}) -
{\f^2 \over 4} \sigma_{mn} F^m_{ij} F^{nij} \} \, ,
\label{eq:10a}
\ee
where $F^m_{ij} \equiv \partial_i A^m_j - \partial_j A^n_i$ ($i,j = 0,1$).
These Abelian gauge fields are not propagating and their contribution is
usually neglected. We propose to take them into account by solving their
equations of motion and writing the corresponding effective potential.
Let us first introduce a very convenient parameterization of the matrix $\sigma_{mn}$:
\be
\sigma_{22} = e^{\eta}\cosh\xi , \,\,\,\, \sigma_{33} = e^{-\eta} \cosh\xi, \,\,\,\,
\sigma_{23} = \sigma_{32} = \sinh\xi \, .
\label{10b}
\ee

After simple derivations (see, e.g., \cite{A1},  \cite{VA}) we exclude the
gauge fields and find the effective potential
\be
V_{\rm eff} = -{1 \over 2 \f^2} \sum_{mn} Q_m (\sigma^{-1})_{mn} Q_n =
-{\cosh \xi\over 2 \f^2} [Q_1^2 e^{-\eta}   - 2 Q_1 Q_2 \tanh \xi
+ Q_2^2 e^{\eta}] \, ,
\label{eq:11a}
\ee
where $Q_m$ are arbitrary constants having pure geometric origin, although they
look like charges of the Abelian gauge fields $F^m_{ij}$. Expressing the
trace in the Lagrangian (\ref{eq:10a}) in terms of the variables $\xi$ and $\eta$,
we derive the Lagrangian in our standard form (\ref{eq:1}):
\be
L = \sqrt{-g} \, \{ \f R(g) + {1 \over 2\f} (\nabla \f)^2  + V_{\rm eff} -
{\f \over 2} [ (\nabla\xi)^2 + (\cosh \xi)^2  \, (\nabla \eta)^2 ] \} \, .
\label{eq:11b}
\ee
This representation is convenient for writing the equations of motion
(\ref{eq:4})-(\ref{eq:6}), for further reductions to dimensions $(1+0)$,
and $(0+1)$, and for analyzing special cases (such as $Q_1 Q_2 =0$, $\xi
\eta =0$). This form is also closer to the original Einstein and Rosen
model, which can be obtained by putting $Q_1 = Q_2 =0$ and $\xi = 0$. It
is also more convenient for analyzing the physical meaning of the
solutions.

The equations of motion (\ref{eq:5}) for the Lagrangian (\ref{eq:11b}) are
\be
2 \f \, \partial_u \partial_v \xi + [\partial_u \f \, \partial_v \xi +
(\partial_u \Leftrightarrow \partial_v)] - 2f \, \partial_{\xi} V_{\rm eff}
- \f \sinh 2\xi \, \partial_u \eta \, \partial_v \eta = 0 ,
\label{eq:11c}
\ee
\be
2 \f \, \partial_u \partial_v \eta + [\partial_u \f \, \partial_v \eta +
2\f \tanh \xi \, \partial_u \xi \, \partial_v \eta  +
 (\partial_u \Leftrightarrow \partial_v)] -
2f (\cosh \xi)^{-2}  \, \partial_{\eta}  V_{\rm eff} = 0 .
\label{eq:11d}
\ee
If $\partial_{\xi} V_{\rm eff} = 0$ and $\partial_{\eta} V_{\rm eff} = 0$,
these equations have solutions with constant $\eta$ and $\xi$ (`scalar
vacuum'). However, for $Q_1 Q_2 \neq0$ we find that the constant solution
of the equations $\partial_{\xi} V_{\rm eff} = 0$, $\partial_{\eta} V_{\rm
eff} = 0$ does not exist because $\xi$ should be infinite:
\bdm
\exp 2\eta = Q_1^2 / Q_2^2 \, ; \,\,\,\,\, \tanh \xi = {\rm sgn (Q_1 Q_2)} ,
\,\,\,\,\, \textrm{i.e.} \,\,\,\,\, \xi = \pm \infty .
\edm
 If $Q_1 Q_2 = 0$, $Q_1^2 + Q_2 \neq 0$,  there exists the constant solution,
  $\xi \equiv 0$, of (\ref{eq:11c})
 while $\partial_{\eta} V_{\rm eff} \neq 0$ and thus there is no constant
 solution of (\ref{eq:11d}). We conclude that
 both $\xi$ and $\eta$ can be constant if  and only if $Q_1 = Q_2 = 0$.
 This agrees with the fact that the flat symmetry (as well as the generalized spherical
 symmetry) does not allow for the existence of `geometric gauge fields' $A_i^m $ (see
next Section). In a more general approach, the generalized spherically symmetric
configurations should be treated as almost spherical axially symmetric solutions,
for which these gauge fields do not vanish. In this sense, the standard spherical
solutions are qualitatively different from the `almost spherical' ones\footnote{
Our approach can be applied \emph{mutatis mutandis} to considering such solutions of
the static axially symmetric theory described by the Ernst equations \cite{Ernst}.
}.

When the potential $V_{\rm eff}$ identically vanishes, Eqs.(\ref{eq:11c}), (\ref{eq:11d})
as well as Eq.(\ref{eq:3}) drastically simplify and we get the Einstein-Rosen
equations for $\xi \equiv 0$. Otherwise we have a nontrivially integrable system of
nonlinear equations belonging to the type considered in \cite{BZ} - \cite{Alekseev}.
With nonvanishing $Q_1$ and/or $Q_2$, even the further reduced (one-dimensional)
equations are nontrivial and it is not quite clear whether they are integrable or
not.

Indeed, let us consider the naive reduction of our theory following simple prescriptions of
\cite{A1}-\cite{A3}. Then the fields and metric coefficients depend on one variable
$\tau = a(u) + b(v)$ and the effective Lagrangian may be written simply as:
\be
L = - {1 \over l}[\dF \dvf  + W {\dvf}^2 - \half \f ({\dxi}^2 + {\deta}^2 \cosh^2 \xi )]
+ l f V_{\rm eff} \, .
\label{eq:a1}
\ee
Here $F = \ln |f|$ and $l = l(\tau)$ is the Lagrange multiplier, which can be expressed
in terms of the metric $g_{ij}$ (but we do not need this expression here).
The equations of motion can be obtained directly from the two-dimensional equations or
from this one-dimensional Lagrangian. Before writing the equations, we absorb $\f$
into $1/l$ and define $\phi \equiv \ln|\f|$. Then the Lagrangian with the new Lagrange
multiplier is
\be
L = - {1 \over l}[\dF \dphi  + \half {\dphi}^2 - \half ({\dxi}^2 + {\deta}^2 \cosh^2 \xi )]
+ l \ve e^{F + \phi} V_{\rm eff} \, ,
\label{eq:a2}
\ee
where $\ve$ is the sign of $f$. After writing the equations of motion we can also
get rid of $l$ by redefining the evolution parameter $\tau$. Note also that we usually
write the equations in terms of the Weyl transformed metric and potential, which are
in our case
$$
\tilde{F} \equiv F + \half \phi \, , \,\,\,\,\,\,\,\,
\tilde{V}_{\rm eff} \equiv e^{-\half \phi} V_{\rm eff} \, .
$$

With arbitrary charges $Q_1$ and $Q_2$, the reduced theory is probably not
integrable. Even when $Q_1 Q_2 =0$, the equations are too complicated to
check their integrability. However, in this case the solution with $\xi \equiv 0$
can be explicitly found. Indeed, in this case we have just one term in the potential
and the theory can be reduced to the N-Liouville integrable model,
which can be explicitly solved \cite{A2}-\cite{A3}.
As was shown in \cite{A2}-\cite{A3}, the
static solutions may have up to two horizons.
A detailed derivations of this solution will be presented elsewhere.

Now consider the well known integrable case $V_{\rm eff} \equiv 0$.
To solve the one-dimensional equations we do not need to use
the inverse scattering method or other advanced
theories. The equations of motion can be directly integrated because they reduce to
the following simple first-order equations (integrals of motion):
\be
2\dF \dphi  + {\dphi}^2 = {\dxi}^2 + {\deta}^2 \cosh^2 \xi \, ; \,\,\,\,\,\,
{\deta} \cosh^2 \xi = c_0 \, ;  \,\,\,\,\,\, {\dxi}^2 = c_1^2 - c_0^2 / \cosh^2 \xi \, ,
\label{eq:a3}
\ee
and, of course, $\dphi = c_{\phi}$, $\dF = c_F$ (obviously,
$2c_{\phi} c_F + c_{\phi}^2 = c_1^2$).
Thus we find
\be
\xi = \ln \{e^{-T_0} H_{\epsilon}(T) +  [1 + e^{-2T_0} H_{\epsilon}^2(T)]^{\half} \} \, ,
\label{eq:a4}
\ee
where $\epsilon$  is the sign of $(c_0^2/c_1^2 - 1)$ and we define
\bdm
H_{-} = \sinh T , \,\,\,\,\,  H_{+} = \cosh T , \,\,\,\,\,
T \equiv c_1 (\tau - \tau_0), \,\,\,\,\, T_0 = -\half \ln |1 - c_0^2/c_1^2| \, .
\edm
By simple integrations we then find $\eta$:
\be
\eta = \eta_0 + \half \ln [1 + e^{-2(T + T_1)}] - \half \ln [1 + e^{-2(T - T_1)}] \, ,
\label{eq:a5}
\ee
where $|c_0 - c_1|/|c_0 + c_1| \equiv e^{-2T_1}$ and $\eta_0$ is a constant.

The expressions for $\xi$, $\eta$, $\phi$, and $F$ have no singularities
in the interval $-\infty < T < +\infty$. A horizon can appear only when $F
\rightarrow -\infty$ while the other three functions are finite. However,
$\phi$ and $\xi$ are infinite for $T \rightarrow \infty$ and thus we have
no horizon. On the other hand, as we mentioned above, the solutions of the
integrable model corresponding to $Q_1 Q_2 = 0$ may have two horizons.
This means that
 the presence of the potential $V_{\rm eff}$ drastically changes the most
important properties of the theory.

In summary of this section, we stress once more that the two-dimensional
theories (\ref{eq:10a}) (and the closely related static axial reductions)
with vanishing gauge fields were extensively used in cosmology (see, e.g.,
\cite{Venezia}, \cite{Lid}) and they are integrable with the aid of modern
mathematical technique (see, e.g., \cite{Alekseev}). However, the
effective potential of the geometric gauge fields most probably destroys
the integrability, even if we further reduce the theory to one dimension.
Nevertheless, the emergence of the potential (\ref{eq:11a}), which under
certain circumstances can imitate effects of the cosmological constant,
may be of significant interest for the present-day cosmology.

\section{Reducing by separating}
The spherical reduction apparently does not allow for appearance of the geometric gauge
fields\footnote{A very careful discussion of the spherically symmetric
space-times and of more general space-times, having subspaces of maximal
symmetry, may be found in \cite{Weinberg} (see also \cite{Petrov}, \cite{Chandra}).}
described in the previous section. Correspondingly, the general spherically symmetric
metric can be written in a simpler form:
\be
ds_4^2 = e^{2\alpha} dr^2 + e^{2\beta} d\Omega^2 (\theta , \phi) -
e^{2\gamma} dt^2 + 2e^{2\delta} dr dt \, ,
\label{eq:8}
\ee
where  $\alpha, \beta, \gamma, \delta$ depend on $(t, r)$ and
$d\Omega^2 (\theta , \phi)$ is the metric on the 2-dimensional
sphere $S^{(2)}$. Substituting this into the action (\ref{eq:7}) and
integrating over the variables $\theta, \phi$ we
find the reduced action\footnote{
This derivation can easily be generalized to any dimension and any number of
the scalar fields with more complex coupling potentials. One can similarly
treat the pseudospherical and flat symmetries as well as any
symmetry given by two Killing vectors. Here $e^{2\beta}$ is the spherical dilaton
denoted by $\f$ in the cylindrical case considered above.
}
with the Lagrangian (\ref{eq:1}), where
\be
U \Rightarrow e^{2\beta} , \,\,\, V \Rightarrow 2 + e^{2\beta}V_4 , \,\,\,
W (\nabla \phi)^2 \Rightarrow 2e^{2\beta} (\nabla \beta)^2 , \,\,\,
Z_{mn} \Rightarrow Z_4 (\psi) e^{2\beta} ,
\label{eq:9}
\ee
and the 2-dimensional metric is given by $e^{2\alpha}, e^{2\gamma}, e^{2\delta}$
(see, e.g., \cite{VA}). Actually, the effective two-dimensional Lagrangian
also contains total derivatives that may be important in some problems but
we will not discuss them here.

The equations of motion for this effective action can easily be derived
and they coincide with Eqs.(\ref{eq:3}) - (\ref{eq:6}) if we pass to the light cone
coordinates. It is not difficult to see (in fact, it is almost evident)
that these equations of motion are identical to the Einstein equations
(see, e.g., \cite{Hajicek})
To simplify the equations, we write them in the limit of the diagonal metric
(formally, one may take the limit $\delta \rightarrow -\infty$). Varying the action
in $\alpha, \beta, \gamma$ and neglecting the $\delta$ - terms
we obtain the Einstein equations for the diagonal components of the Einstein
tensor. Varying the action in $\psi$ we find the equation for $\psi$. Finally,
by varying in $\delta$ we find one more equation corresponding to
the non diagonal component of the Einstein tensor; it is not a consequence of
other equations and is a combination of the two constraints (\ref{eq:4}).

The simplest way to write all necessary equations is to write the
2-dimensional effective action in the coordinates (\ref{eq:8}). First making
variations in $\delta$ we find (in the limit $\delta \rightarrow -\infty$)
the constraint
\be
 {\dot{\beta}}^{\prime} + \dot{\beta} {\beta}^{\prime} -
   \dot{\beta} {\gamma}^{\prime} - \dot{\alpha} {\beta}^{\prime} \,\,
  = \,\, \half Z_4 \dot{\psi} {\psi}^{\prime} ,
\label{eq:10}
\ee
where $\dot{\psi} \equiv \p_t \psi \,$, ${\psi}^{\prime} \equiv \p_r \psi \,$, etc.
The other equations can be derived (in the diagonal limit) from the
effective Lagrangian
\be
L_{\rm eff} \,\, = \,\, V_{\rm eff} + L_{\rm t} + L_{\rm r}  ,
\label{eq:11}
\ee
where we omitted the $\delta$-dependence and total derivative terms.
The sum of the `$r$-Lagrangian'
\be
L_{\rm r} \,\, = \,\, e^{ -\alpha + 2\beta + \gamma} (2 {{\beta}^{\prime}}^2
+ 2{\beta}^{\prime}{\gamma}^{\prime} + Z_4 {{\psi}^{\prime}}^2)  ,
\label{eq:12}
\ee
with the `$t$-Lagrangian'
\be
L_{\rm t} \,\, = \,\,  - e^{\alpha + 2\beta - \gamma} (2 {\dot{\beta}}^2
+ 2{\dot{\beta}} \dot{\gamma} + Z_4 {\dot{\psi}}^2)  ,
\label{eq:13}
\ee
as well as the constraint (\ref{eq:10}) are invariant under the substitution
$\p_r \Leftrightarrow i\p_t$ and $\alpha \Leftrightarrow \gamma$. This means that the
equation of motion are invariant under this transformation, as the effective
potential\footnote{
Here, in addition to the case of the spherical symmetry
($k=1$) we include the cases of pseudospherical ($k=-1$) and flat ($k=0$) symmetries.
},
\be
V_{\rm eff} \, = \, V_4 \, e^{ \alpha + 2\beta + \gamma} + 2k \, e^{\alpha + \gamma} ,
 \,\,\,\,
 k = 0, \pm1 ,
\label{eq:14}
\ee
is naturally invariant. At first sight, this invariance may look trivial but one
should recall that in higher dimensions there is no complete symmetry between
space and time. Thus the simple relation between static and cosmological solutions
suggested by this symmetry may give some new insight into both classes of objects.
Even apart from any physical interpretation, this symmetry
allows us to economize writing equations and it is extremely useful in
considering separation of variables outlined below. In particular,
these transformations allow us to derive cosmological solutions  corresponding
to static (black hole) solutions and vice versa. Although this is a special case
of the formulated duality relation we call it `static-cosmological' (SC) duality.

To illustrate how the separation of the variables looks like we write the three
remaining equations (in addition to Eq.~(\ref{eq:9})):
\be
[2e^{-2\alpha} ({\beta}^{\prime \prime} + 2{{\beta}^{\prime}}^2 -
{\beta}^{\prime} {\alpha}^{\prime} + {\beta}^{\prime} {\gamma}^{\prime})] -
[\alpha \Leftrightarrow \gamma , \,\, {\p}_r \Rightarrow {\p}_t] \, =
 \, V_{\rm eff} \,  e^{-\alpha - 2\beta - \gamma} \, ,
 \label{eq:15}
\ee
\be
[2e^{-2\alpha} ({\beta}^{\prime \prime} + {{\beta}^{\prime}}^2 -
{\beta}^{\prime} {\alpha}^{\prime} - {\beta}^{\prime} {\gamma}^{\prime})] +
[\alpha \Leftrightarrow \gamma , \,\, {\p}_r \Rightarrow {\p}_t] \, = \, Z_4 E_+ \, ,
\label{eq:16}
\ee
where we denote
\be
E_{\pm} \, \equiv \, e^{-2\alpha} {{\psi}^{\prime}}^2 \, \pm \,
e^{-2\gamma} {\dot{\psi}}^2  .
\label{eq:17}
\ee
The third equation has a similar structure
\be
[e^{-2\alpha} ({\gamma}^{\prime \prime} + {{\gamma}^{\prime}}^2 -
{\gamma}^{\prime} {\alpha}^{\prime} - {{\beta}^{\prime}}^2 ] -
[\alpha \Leftrightarrow \gamma , \,\, {\p}_r \Rightarrow {\p}_t] \, = \,
- k e^{-2\beta} + \half Z_4 E_- \, ,
\label{eq:18}
\ee
We see that the equations are duality invariant.
In practical procedures of the separation it may be convenient to also
use an additional (dependent) equation for $\psi$ and work with some
linear combinations of the written equations.

To make a separation of the space and the time variables possible we should
try to write all the equations in the form
\be
\sum_{n=1}^{N} T_n(t) R_n(r) = 0 ,
\label{eq:18a}
\ee
where $T_n$ depends only on functions (and their derivatives) of the time variable,
while $R_n$ depends only on space functions. Then, dividing by one of the functions
and differentiating w.r.t. $r$ or $t$ we finally find equations for functions of
one variable depending on constants, which functionally depend on functions of
the other variable\footnote{This is one of several possible
approaches to solving these equations. We may call it `dividing and differentiating'
or simply `dd-procedure'.}.
For $N=2$ this is obvious: $T_1/T_2 = R_1/R_2 = C$.
For $N=3$ we may write, for instance,
\bdm
(T_1/T_3)(R_1/R_3) + (T_2/T_3)(R_2/R_3) + 1 = 0
\edm
and then differentiate this equation w.r.t. $r$ or $t$, thus reducing the equation to
the $N=2$ case with a new arbitrary constant appearing due to differentiations.

It is evident that, to write the equations in the form (\ref{eq:18a}), we should
make some \emph{Ansatz} allowing us to write all the terms as products of functions
of one variable.
It is clear that to separate the variables $r$ and $t$ in the metric we should
require that
\be
\alpha = \alpha_0(t) + \alpha_1(r) , \,\,\, \beta = \beta_0(t) + \beta_1(r) , \,\,\,
\gamma = \gamma_0(t) + \gamma_1(r) , \,
\label{eq:19}
\ee
Then, the potentials $V_4$ and $Z_4$ must be either constant or have
the necessary multiplicative form.
Depending on the analytic form of the potentials, this is possible in two
principal cases
\be
\psi = \psi_0(t)  + \psi_1(r) , \,\,\,\, {\rm (a)} \,  \,\,\,\, {\rm or} \,\,\,\,
      \psi = \psi_0(t) \psi_1(r) , \,\,\,\,  {\rm (b)} \, .
\label{eq:20}
\ee
Here we will not try to find and classify all possible cases of separation and
mention only typical ones. If $\dot{\psi} \psi^{\prime} = 0$, a separation
is possible for generic potentials. If $\dot{\psi} \psi^{\prime} \neq 0$,
there are three obvious classes of the potentials that allow the
separation: 1.~constant potentials $V_4$ and $Z_4$; 2.~exponential
$V_4(\psi)$ and $Z_4(\psi)$
(with the \emph{Ansatz} (\ref{eq:20}a); 3.~power dependent $V_4(\psi)$ and
$Z_4(\psi)$ (with the \emph{Ansatz} (\ref{eq:20}b). Note that the case of the
constant $Z_4$ and exponential $V_4$ is often met in dimensional reductions
of gravity and supergravity.

In four-dimensional theories obtained by the chains of dimensional reductions
discussed in Introduction, the potentials are exponentials of linear sums
of the $\psi$-fields. A rather general theory, slightly more general than
(\ref{eq:7}),
\be
S_4 = \int d^4 x \, \sqrt{-g_4} \, [ R_4 + V_4 (\psi) +
\sum_{n=1}^N Z^{(n)}_4 (\psi) (\nabla \psi_n )^2] ,
\label{eq:20a}
\ee
depends on $N$ scalar matter fields $\psi_n$. The potential $V_4$ is a sum of
linear exponentials of the fields $\psi_n$,
\be
V_4 = \sum_{k=1}^K g_k \exp \, [\sum_{m=1}^N \psi_m a_{mn}] \, ,
\label{eq:20b}
\ee
and $Z^{(n)}_4$ are either constants or simple linear exponentials of some of the fields $\psi$
(see, e.g., \cite{Lid} and \cite{VA})\footnote{The cylindrical theory and the static
axial theory also belong to this class.
}.
With the additive \emph{Ansatz} for $\psi$,
\be
\psi_n = \psi_{0n}(t)  + \psi_{1n}(r) ,
\label{eq:20c}
\ee
the separation is also possible for the potential and $Z$ terms.
To simplify the presentation we discuss here the case of one scalar field
with the additive \emph{Ansatz} (the generalization to $N$ fields with the same
\emph{Ansatz} is not difficult).

Inserting \emph{Ansatzes} (\ref{eq:19}) - (\ref{eq:20}) into the equations
(\ref{eq:15}), (\ref{eq:16}), (\ref{eq:18}) we can find further conditions for
the separation (when all the equations can be rewritten in the form of Eq.(\ref{eq:18a})).
To simplify the discussion we choose\footnote{Sometimes,
this may be inconvenient. For example, for the standard form of the Schwarzschild solution
$2\alpha_1 = -\ln(1 - r_0 / r)$.
}: $\alpha_1 = 0$ and $\gamma_0 = 0$.
Note that this does not significantly restrict our local considerations
because this is equivalent
to changing the coordinates $(t,r)$ to $(\bar{t}, \bar{r})$:
$$
\bar{t} = \int dt \, e^{\gamma_0(t)}, \,\,\,\,\,\,\,\,
\bar{r} = \int dr \, e^{\alpha_1(r)} \, .
$$
Then, analyzing (\ref{eq:15}) - (\ref{eq:18}), we see that all the terms, except
$ke^{2\beta}$ (for $k \neq 0$!), will have the form $T_n(t) R_n(r)$
after multiplying the equations by
$e^{2\alpha_0} e^{2\gamma_1}$. By applying the dd-procedure it is easy to prove that
$ke^{2\beta}$ can be presented in the required form if at least one of the two
(dual) conditions,
\be
 \textrm{I.} \,\,\,\, \dot{\alpha_0} \, = \, \dot{\beta_0}  \,\,\,\,\,\, \textrm{or}
 \,\,\,\,\,\, \textrm{II.} \,\,\,\, {\beta_1}^{\prime} \, = \, {\gamma_1}^{\prime} \, ,
\label{eq:20d}
\ee
is satisfied for $k \neq 0$. When $k=0$, this restriction is unnecessary.

The second strong restriction on the separated metric functions
is given by the constraint (\ref{eq:10}).
 Let us first consider the case
$\dot{\psi} \psi^{\prime} = 0$. Then it is not difficult to prove that its
solutions are:
\be
{\rm 1}. \,\, \dot{\alpha_0} \, = \, \dot{\beta_0}  = 0 \, ; \,\,\,\,\,\,\,\,
{\rm 2}. \,\, {\beta_1}^{\prime} \, = \, {\gamma_1}^{\prime} = 0 \, ;
\label{eq:21}
\ee
\be
{\rm 3}. \,\, \dot{\alpha_0} \, = 0 \, , \,\,\,\,
{\beta_1}^{\prime} \, = \, {\gamma_1}^{\prime} \, ; \,\,\,\,
{\rm 4}. \,\, {\gamma_1}^{\prime} = 0 \, , \,\,\,\,
\dot{\alpha_0} \, = \, \dot{\beta_0} \, ; \,\,\,\,
{\rm 5}. \,\, \dot{\beta_0} \, = \, {\beta_1}^{\prime} = 0 \, .
\label{eq:21a}
\ee
When no one of this conditions is satisfied, (\ref{eq:10}) can be written as
\be
 \,\,\, {\gamma_1}^{\prime} / {\beta_1}^{\prime} \, +
 \dot{\alpha_0} / \dot{\beta_0} \, = 1 \, ,
\label{eq:22}
\ee
 and the solution of this equation is obviously ($-\infty < C < +\infty$)
\be
 {\rm 6} . \,\,\, {\gamma_1}^{\prime} = (1-C) {\beta_1}^{\prime} \, , \,\,\,\,\,\,
 \dot{\alpha_0} = C\dot{\beta_0} \, .
\label{eq:22a}
\ee
The conditions 2 and 4 are dual to the conditions 1 and 3, respectively.
The conditions 5 and 6 are obviously self-dual, the conditions 3 and 4
follow from (\ref{eq:22a}) for $C=0$ and $C=1$, respectively. If we take into account
the condition (\ref{eq:20d}), we find that for $k \neq 0$ there are
four basic configurations: A.~$\dot{\alpha_0} = \dot{\beta_0} \, = 0$ or the dual;
B.~$\dot{\alpha_0} = \dot{\beta_0}, \, {\gamma_1}^{\prime} = 0$
or the dual. The A-configuration corresponds to naive static or cosmological reductions.
The B-configuration cannot be obtained by naive reductions. With $k = 0$, we simply use
the conditions (\ref{eq:21}) - (\ref{eq:22a}).

One can similarly treat the case $\dot{\psi} \psi^{\prime} \neq 0$.
To simplify the discussion let us suppose that $Z_4 = -2$.
Applying the dd-procedure to the constraint (\ref{eq:10}) written as
\be
{\dot{\alpha_0} \over \dot{\psi_0}} \, {{\beta_1}^{\prime} \over {\psi_1}^{\prime}} \,
  + \, {\dot{\beta_0} \over \dot{\psi_0}} \,
 {{{\gamma_1}^{\prime} \, - \, {\beta_1}^{\prime}} \over {\psi_1}^{\prime}}
 \, - \, 1 = 0 \, ,
\label{eq:22b}
\ee
we can prove that its solutions are
\be
\dot{\alpha_0} \, = \, (1-C) \dot{\beta_0} \, , \,\,\,\,\,
 \dot{\psi_0} \, = \, C_1 \dot{\beta_0} \, , \,\,\,\,\,
 {\psi_1}^{\prime} \, = \, C_1^{-1} ({\gamma_1}^{\prime} - C {\beta_1}^{\prime}) \, ,
\label{eq:22c}
\ee
\be
\dot{\beta_0} \, = \, 0 \, , \,\,\,\,\,
 \dot{\psi_0} \, = \, C_1^{-1} \dot{\alpha_0} \, , \,\,\,\,\,
 {\psi_1}^{\prime} \, = \, C_1 {\beta_1}^{\prime} \, ,
\label{eq:22d}
\ee
and two solutions that are dual to these.

Now, if the potentials $V_4$ and $Z_4$ are separable and the constraint (\ref{eq:10})
as well as one of the constraints (\ref{eq:20d}) (for $k \neq 0$) are satisfied,
all the equations of motion can be written in the separated form (\ref{eq:18a}).
If there are several scalar fields $\psi_n$ we should add the equations obtained
by varying the Lagrangian also w.r.t. these fields. Following this way, we can obtain
all the standard black holes and cosmologies and, in addition, other spherically
symmetric static and cosmological solutions related by the SC-duality.
It is important to emphasize that, when we have many scalar matter fields
and rather complex potentials, the equations of motion are, in general, not integrable.
Note also that the
distinction between the cosmological and static solutions is not trivial because general
reduced solutions depend on both $r$ and $t$. Nevertheless, we should regard the procedure
of separation as a dimensional reduction. The meaning of this can be clarified
by the following examples.

Let us first choose
condition 2 of (\ref{eq:21}), which says that $\beta_1$ and
$\gamma_1$ are constant\footnote{A cosmological model with this metric
was studied in \cite{Komp}, \cite{Kant}.}
 (without loss of generality we may suppose that they vanish).
Then  the equations of motion (\ref{eq:15}) - (\ref{eq:18})
for the remaining dynamical functions
$\alpha_0$, $\beta_0$, $\psi_0$,
\be
(\ddot{\beta}_0 + 2 {\dot{\beta_0}}^2 + \dot{\beta_0} \dot{\alpha_0})  +
k e^{-2\beta_0} \, =  \, -\half V_4(\psi_0) \, ,
\label{eq:23}
\ee
\be
(\ddot{\beta}_0 + {\dot{\beta_0}}^2 - \dot{\beta_0} \dot{\alpha_0})
 \, =  \, \half Z_4 \, {\dot{\psi_0}}^2   \, ,
\label{eq:23a}
\ee
\be
(\ddot{\alpha}_0 + {\dot{\alpha_0}}^2 - \dot{\beta_0}^2)  -
k e^{-2\beta_0} \, =  \, \half Z_4 \, {\dot{\psi_0}}^2 \, ,
\label{eq:23b}
\ee
define a cosmological model.
 From (\ref{eq:23})  and (\ref{eq:23a}) we derive
the integral of motion, which is the total (gravity plus matter) energy of the system
\be
{\dot{\beta_0}}^2 + 2 \dot{\beta_0} \dot{\alpha_0}  + k e^{-2\beta_0} +
\half V_4(\psi_0) + \half Z_4 \, {\dot{\psi_0}}^2  \, = 0.
\label{eq:23c}
\ee
Thus we find the complete system of equations
for $\alpha_0$, $\beta_0$, $\psi_0$
(recall that we take the coordinates in which $\alpha_1 = \gamma_0 = 0$).
Of course, one can see that we recovered the naive cosmological reduction.
This cosmology does not coincide with the standard FRW cosmology, which can be obtained
by using the conditions ${\gamma_1}^{\prime} = 0$,
 $\dot{\alpha_0} = \dot{\beta_0}$, and $\psi^{prime}_0 = 0$. The naive cosmology is
 unisotropic because $R_{11}^{(3)} =0$ while $R_{22}^{(3)} = k$, where $R_{ij}^{(3)}$
is the Ricci tensor of the 3-dimensional subspace $(r, \theta, \phi)$.
For the naive cosmology, the Ricci curvature of the 3-space is simply
$R^{(3)} = 2ke^{-2\beta_0(t)}$.

The naive static reduction may be reproduced simply by using our SC-duality.
The Schwarzschild black hole then can be obtained if we take $V_4 =0$ and $Z_4 =0$.
Otherwise we have a static state of gravitating scalar matter. For the generic functions
$V_4(\psi)$ and $Z_4(\psi)$ the equations of motion for both dual theories are not integrable
but, if the static theory has horizons (this does not contradict to the `no hair' theorem,
if $V_4$ depends on $\psi$, see \cite{A1}-\cite{A3}), we may construct analytic
perturbation theory near each horizon (see \cite{FM}). It would be interesting
to construct a `dual' perturbation theory for the cosmological solutions.
Note that the perturbation theory can be applied not only to the naive reductions.

With less restrictive \emph{Ansatzes} we may construct other cosmological models and
static configurations that are dual to them. For example, with
the condition (\ref{eq:22a}), we find in the equations of motion the terms
depending both on $r$ and on $t$. Thus the more general procedure of separation
should be applied.
The interpretation of the solutions as cosmological or static
requires more care and will be discussed in a separate publication, where further
examples will also be presented. Constructing the one-dimensional Lagrangians
producing the reduced equations of motion requires more care also.
Especially, we should not completely fix the gauge. Even in the naive reduction
(\ref{eq:23}) - (\ref{eq:23c}) we must avoid choosing the obvious gauge
$\gamma_0 = 0$ because $e^{\gamma_0}$ plays the role of the Lagrange multiplier
$l(\tau)$ in the one-dimensional Lagrangian (\ref{eq:a1}).

It is interesting that there
may exist some `intermediate' cases that are more symmetric under the duality
transformation being neither static nor cosmological. It would be premature to call
them `self - dual' before a detailed study of them will be undertaken.

Here we considered the separating of variables approach for the spherically
reduced gravity. With due care, it can be applied to the generalized cylindrical
theory (\ref{eq:11b}). It is of interest to apply separating to reducing static
axial theory with KMKF potentials. This may allow us to describe essentially generalized
perturbed spherical states that may be considered as more realistic models of black holes.

\section{Conclusion}
In a separate publication I will present a complete list of all mentioned
reductions and their relation to black holes, cosmologies, and
waves (especially, the cylindrical ones). I will also make an attempt to compare the
models obtained by the approach of this paper to known cosmological and
static solutions derived by other methods  and classified by their
group theoretical properties. At the moment, relations of our dynamical
classification to the group theoretical ones is not clear.

An interesting new topic is dimensional reduction to waves.
Here I only mention the wave - like solution obtained in the integrable model of
(1+1)-dimensional gravity coupled to $N$ scalar matter
fields \cite{VA}, \cite{A3}.
The general solution of the model depends on the chiral moduli fields
$\xi_n(u)$, $\eta_n(v)$ that move on the surfaces of the spheres $S^{(N)}$.
The naive reduction to one-dimensional theories emerges when
the moduli fields are constant and equal, $\xi_n = \eta_n$. When they are
constant but otherwise arbitrary, we have a new class of reduced solutions
that correspond to waves of scalar matter coupled to gravity.
Under certain conditions, these waves may be localized in space and time
and thus may be regarded as a sort of solitary gravitational waves\footnote{
A special solution of this kind has recently been found in \cite{A3} and will be
generalized and discussed in more detail in the forthcoming paper \cite{VA1}.
Note that our solitary waves do not seem to have a relation to possible
soliton - like states in the theories with the `sigma - model' - like
coupling of the scalar fields to gravity,
which can be obtained from (\ref{eq:11b}) with $V_{\rm eff} \equiv 0$
(they are studied in \cite{BZ} - \cite{Alekseev},
see also a discussion in \cite{A3} as well as a simplified explicitly soluble model
of scalar waves in dilaton gravity proposed in \cite{FI}).
}.
The very origin of these waves signals existence of a close relation between main gravitational
objects - black holes, cosmologies and waves. This relation
was studied in some detail for static states and cosmologies and so was called
the static-cosmological duality. In the integrable models,
transitions between static and cosmological states are possible and,
moreover, the waves play a significant role in these transitions. This observation,
which does not actually require integrability, may open a way to studies of real
physical connections between these apparently diverse objects.

In summary, one may identify at least three types of dimensional reduction:
the `standard' or `naive' reduction which supposes that functions of two variables
depend on one variable only, the reduction by separating the variables, and
the reduction in moduli spaces supposing that the moduli functions become
constants. In all cases the important problem is to find the Lagrangians and
Hamiltonians for the reduced systems. This is not difficult for  naive reductions
and for simple reductions based on separating. It is not clear how
to do this with the last, so to speak, `moduli reduction`. In addition, it is not
clear how to do such a reduction for not integrable systems.

Finally, we wish to emphasize once more that the `geometric potentials' can
emerge in `almost spherical' (perturbed axial) states. It must be not very difficult
to apply our considerations to such states and we hope to do this in near future.


 \section{Acknowledgments} I very much appreciate support of
 the Department of Theoretical Physics of the University of Turin and
 INFN (Section of Turin) as well as of the MPI and the W.~Heisenberg
 Institute (Munich), where some results were obtained.
 I am especially grateful to V.~de Alfaro for his support for many
 years and very fruitful collaboration.

 This work was supported in part by the Russian Foundation for Basic
 Research (Grant No. 06-01-00627-a).

\end{document}